# Modelling viscoelastic adhesion and friction in sliding contact mechanics

C.Mandriota[1,*] , N. Menga[1] and G. Carbone[1].

[1]Department of Mechanics, Mathematics and Management, Polytechnic University of Bari , Italy
*`cosimo.mandriota@poliba.it`

**Abstract.** We present our recent study on rough adhesive contacts of viscoelastic materials in steady-state sliding, focusing on the interplay between adhesion and viscoelasticity by means of a novel energy approach. We investigate tribological features over a wide range of velocity values, exploring the effect of small- and large- scale viscoelasticity on the overall contact behavior. The former is associated with viscoelastic dissipation close to the adhesive neck at the contact edges; the latter refers to material hysteresis involving the entire bulk of the solid. Depending on the sliding velocity, we predict highly enhanced adhesive strength compared to the purely elastic conditions, increased friction compared to the adhesiveless case, and non-monotonic trend of the energy release rates at the contact leading and trailing edges. Most of our results are supported by existing experiments.

## 1    Introduction

Understanding and predicting adhesion and friction in rolling or sliding contact between polymeric materials is a crucial issue in several engineering applications (bio-inspired adhesives, Micro-Electro-Mechanical-Systems, micro grippers, protective coatings, tire-road contact, pressure sensitive adhesives, lubrication, wear, windscreen wipers, adhesive suits and gloves, structural adhesives). Nevertheless, most of existing theoretical studies can predict the contact behavior only by neglecting adhesion [1-6] or viscoelasticity [7-13] and a comprehensive theory of viscoelastic adhesion is still lacking. On the other hand, experimental data from the literature highlight that the coupling between adhesion and viscoelasticity might strongly affect the tribological features of the contact, either in terms of effective adhesive strength and friction, making the development of a comprehensive theory of viscoelastic adhesion a fundamental target in contact mechanics research.

Experimental studies from the literature [14-16] offer a wide investigation of enhanced adhesion induced by the so-called small-scale viscoelasticity, which is found to play a major role up to very low sliding/rolling velocity values [17]. Despite in such conditions the bulk of the material almost behaves as a purely elastic solid, a large amount of viscoelastic dissipation is locally induced by the adhesive neck of the deformed profile. This phenomenon eventually results into asymmetric contact shape, with contact area highly enlarged at the trailing edge of the profile, where the local behavior resembles viscoelastic fracture with increased energy release rate [18,19]. Experiments [14] demonstrate that these phenomena make the adhesive interface able to withstand significantly larger tractive loads compared to the static (i.e., purely elastic) case.

Similarly, the effect of the interplay between adhesion and viscoelasticity on friction was investigated in [20,21]. Reported data evidence that adhesion strongly enlarges friction with respect to adhesiveless conditions, and suggest that the overall response is affected by complex phenomena of coupling between small- and large- scale hysteresis. Indeed, besides the overmentioned local hysteresis, adhesion might also increase the contact area and in turn induce additional amount of viscoelastic dissipation involving the whole bulk of the solid (i.e., large-scale viscoelasticity). Various theories and numerical studies [1-6,22] can quantify the portion of friction attributable to bulk viscoelasticity in adhesiveless conditions, but the accurate prediction of the adhesive effect on the overall viscoelastic friction has been only recently presented [23,24].

## 2   Formulation

We consider a linear viscoelastic half-space sliding in adhesive steady contact with a rigid indenter. Assuming a reference frame co-moving with the indenter, the stationary surface displacement $u(x)$ and the pressure field $p(x)$ are related each other through the convolution product:

$$u(x) = \int d^2x_1 G_v(x - x_1, V) p(x_1) \qquad (1)$$

where $G_v(x, V)$ is the viscoelastic Green's function, parametrically depending on the sliding velocity $V$. Eq. (1) enables the calculation of pressure and displacement fields for given contact domain. However, an additional equation is needed since the contact domain is unknow. At this aim, we enforce the energy balance at virtual quasi-static perturbation of the contact area $A$:

$$dL = \Delta\gamma \delta A \qquad (2)$$

where $\delta A$ is the infinitesimal variation of the contact area, $\Delta\gamma$ is the Douprè work of adhesion [7-13] and $dL$ is the work of internal stresses. In the purely elastic case, $dL$ equates the variation of elastic energy $U$, leading to the Griffith equation $G = \Delta\gamma$, where the energy release rate $G = \partial U/\partial A$. When viscoelasticity is involved, the internal work must include a non-conservative (i.e. path-dependent) work contribution:

$$\delta L = dU + \delta L_P \qquad (3)$$

Following [23], $\delta L_P$ can be calculated as function of the contact pressure distribution:

$$\delta L_P = - \int d^2x d^2x_1 G_V^O(x - x_1, V) p_+(x) p_-(x_1) \qquad (4)$$

with $p_-(x) = p(x, A)$ and $p_+(x) = p(x, A + \delta A)$ being the contact pressure fields developed within the contact area $A$ and increased contact area $A + \delta A$ respectively and $G_V^O(x) = \frac{1}{2}\big(G_v(x) - G_v(-x)\big)$ is the odd part of the viscoelastic Green's function, vanishing in the purely elastic case [7]. The elastic energy of the system is:

$$U = \frac{1}{2} \int d^2x p(x) u(x) \qquad (5)$$

The effect of viscoelasticity on the energy balance is better emphasized by combining Eqs. (2) and (3), and observing that the term $\delta L_P$ depends on the sliding velocity value and can be either positive or negative, resulting in a velocity dependent value of the energy release rate, increased or decreased with respect to the Douprè work of adhesion:

$$G(V) = \Delta \gamma - \frac{\delta L_P}{\delta A}(V) \qquad (6)$$

## 3   Results and discussion for a wavy substrate

We study the case of rigid sinusoidal indenter of wavelength $\lambda$ and height $\Lambda$. Notably, the surface displacement is prescribed within the contact area, where Eq. (1) can be written as:

$$\Delta + \Lambda(\cos(kx) - 1) = \int_{-x_1}^{x_2} G_v(x - s) p(s) ds, \qquad -x_1 < x < x_2 \qquad (7)$$

with $\Delta$ being the contact penetration, $k = 2\pi/\lambda$ and $x_1, x_2$ the coordinates identifying the asymmetric contact area (see Fig. 1). According to [1,2], the periodic viscoelastic Green's function is:

$$G_V(x) = J(0^+)\vartheta(x) + \int_{0^+}^{+\infty} \vartheta(x + Vt) \dot{J}(t) dt \qquad (8)$$

Where $\vartheta(x)$ is the elastic periodic Green's function:

$$\vartheta(x) = -\frac{2(1-v^2)}{\pi} \log\left[2\left|\sin\left(\frac{kx}{2}\right)\right|\right] \qquad (9)$$

And $J(t)$ the viscoelastic Creep's function:

$$J(t) = H(t)\left[\frac{1}{E_0} - \left(\frac{1}{E_0} - \frac{1}{E_\infty}\right) \exp(-t/\tau)\right] \qquad (10)$$

Here, $E_0$ and $E_\infty$ are the low and high frequency elastic moduli of the material, respectively. Notably, the pressure field $p(x)$ vanishes out of contact, thus Eq. (7) can be numerically solved for the pressure distribution $p(x)$ as function of the contact coordinates $x_1$ and $x_2$ for given contact penetration. The adhesive stress singularity is properly computed by relying on the numerical procedure addressed in [7], based on adaptative non-uniform mesh. Once $p(x)$ is determined, the displacement field $u(x)$ can be calculated over the entire wavelength through Eq. (1). The equilibrium values of contact coordinates are found by enforcing Eq. (6) at the contact trailing and leading edges considering independent infinitesimal positive perturbations $\delta x_1$ and $\delta x_2$, yielding to energy release rates $G_1$ (trailing edge) and $G_2$ (leading edge):

$$G_1(V) = \frac{\partial U}{\partial x_1} = \Delta\gamma - \frac{\delta L_{P,1}}{\delta x_1}(V) \qquad (11)$$

$$G_2(V) = \frac{\partial U}{\partial x_2} = \Delta\gamma - \frac{\delta L_{P,2}}{\delta x_2}(V) \qquad (12)$$

Results reflect scale-dependent phenomena ascribable to the interplay between adhesion and viscoelasticity, as clearly suggested by Fig. 1, depicting predicted contact configurations with increasing sliding velocity under constant remotely applied pressure $p_\infty$. At quite low velocity (i.e., Fig.1a), the adhesive behavior is governed by the small-scale viscoelasticity and the contact is found highly asymmetric and enlarged at the trailing edge, in agreement with experimental observation [14-16]. In such conditions, the bulk material almost behaves elastically as excited at very low frequencies, with main frequency roughly estimable as $\sim \frac{V}{\lambda} \ll \frac{1}{\tau}$, and the majority of viscoelastic dissipation occurs close to the contact trailing edge, where the local frequency of excitation $\sim V/R$, with $R$ being the local radius of curvature of the deformed profile (notably, $R \ll \lambda$). This eventually results into increased effective energy of adhesion, i.e. contact area and pull-off load enlargement with respect to static conditions (see also Figs. 2, 3). When the velocity is increased (i.e., $R/\tau < V < \lambda/\tau$, referring to Fig. 2b) the contact region gradually moves forward with respect to the indenter's summit. Under these conditions, scale separation no longer holds as viscoelastic losses also involve the bulk, yielding to overall contact behavior affected by coupling between small- and large-scale viscoelasticity. At even higher velocity (fig. c), i.e. when $V \sim \frac{\lambda}{\tau} \gg \frac{R}{\tau}$, the contact is governed by bulk viscoelasticity, leading to contact area completely shifted forward, as predicted in the adhesiveless viscoelastic contact case [1,2]. Notably, purely elastic contact conditions are recovered at extremely low and high velocity values, corresponding to asymptotic elastic response with moduli $E_0$ and $E_\infty$ respectively, occurring everywhere within the viscoelastic solid. In this scenario, the non-conservative work contributions vanish and symmetric contact solutions (i.e., $x_1=x_2$) are recovered with $G_1 = G_2 = \Delta\gamma$, also yielding to vanishing friction (see Figs. 3, 4).

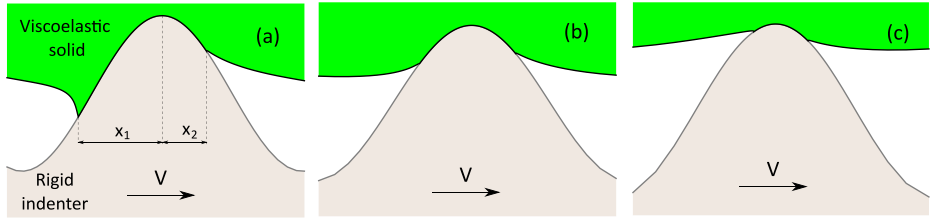

**Fig. 1.** The predicted contact configurations with sliding velocity $V$ increasing from (a) to (c) under load controlled conditions.

The viscoelastic-induced enhanced adhesion is more deeply investigated in Fig. 2, reporting the dimensionless half-width of contact $ka$ and the dimensionless pull-off remote pressure $|p_{out}|$ as function of the dimensionless sliding velocity $kV\tau$, both exhibiting non-monotonic trends. Similar increased adhesive performances have been observed in experiments conducted on rigid cylinders rolling in adhesive contact with rubber substrates, reporting increasing trend of the contact area vs. rolling velocity [15] and highly enlarged pull-off force observed as long as rolling occurs [14]. We predict that this phenomenon is maximum at intermediate velocity, where the contact width enlargement is significative either with respect to the zero velocity value, and compared to adhesiveless conditions (also reported with dashed line). At high velocity values, the viscoelastic bulk stiffening leads to high pull-off remote pressure and low contact area values. Moreover, the overall non-monotonic trend of the pull-off load suggests that the increase of adhesive strength is not only ascribable to the bulk stiffening but is also significantly affected by the small-scale viscoelastic hysteresis.

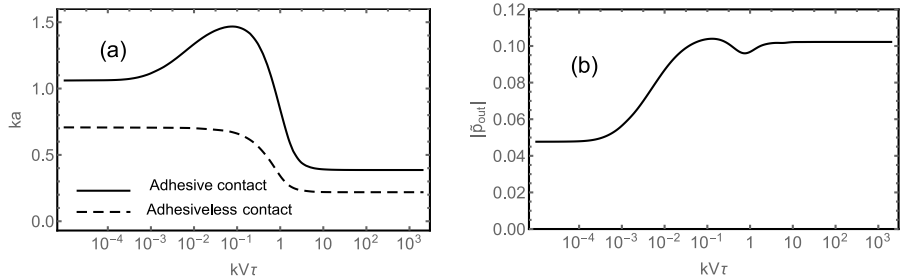

**Fig. 2.** (a) The dimensionless half-length of contact $ka$ shown as function of the dimensionless sliding velocity $kV\tau$, either in adhesive (solid line) and adhesiveless (dashed line) conditions at constant remotely applied dimensionless pressure $\tilde{p}_\infty = \frac{2p_\infty}{E_0^* k\Lambda} = 0.12$. (b) The absolute value of the dimensionless pull-off remote pressure $\tilde{p}_{out}$ shown as function of the dimensionless sliding velocity $kV\tau$. Results are shown for dimensionless energy of adhesion $\tilde{\gamma} = \frac{k\Delta\gamma}{\pi E_0^*} = 0.008$ and $\frac{E_\infty}{E_0} = 10$.

Fig.3 reports the trend of the reduced energy release rates $G_1/\Delta\gamma$ (trailing edge) and $G_2/\Delta\gamma$ (leading edge) shown as function of the dimensionless sliding velocity. At low velocity values the local crack behavior resembles independent closing and opening cracks, yielding to $G_1/\Delta\gamma$ and $G_2/\Delta\gamma$ respectively increasing and decreasing with the sliding velocity, in agreement with predictions of theories of viscoelastic crack propagation and healing [18,19,25]. Anyway, this behavior is limited to the small-scale viscoelastic regime and the effect of bulk viscoelasticity makes the overall trend non-monotonic. According to existing studies [26], this effect is ascribable to the presence of finite lengths in the system, in our case the length of the contact region. Interestingly, at quite high velocity, i.e. when $x_2 > x_1$ (referring to fig. 1 (c)), we observe $G_1 < \Delta\gamma$ and $G_2 > \Delta\gamma$, indicating that the non-conservative work changes sign under these conditions (see Eqs. (11) and (12)).

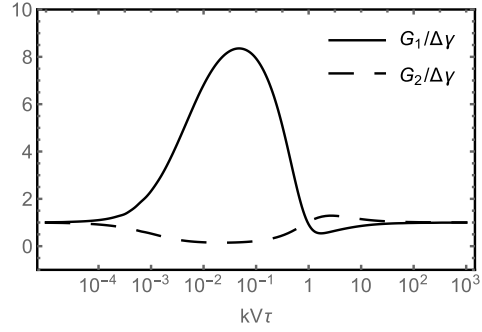

**Fig. 3.** The reduced energy release rate $G/\Delta\gamma$ as function of the dimensionless sliding velocity $kV\tau$ at constant remotely applied dimensionless pressure $\tilde{p}_\infty = \frac{2p_\infty}{E_0^* k\Lambda} = 0.02$. Solid line refers to the contact trailing edge ($G_1/\Delta\gamma$), dashed line refers to the contact leading edge ($G_2/\Delta\gamma$). Results are shown for dimensionless energy of adhesion $\tilde{\gamma} = \frac{k\Delta\gamma}{\pi E_0^*} = 0.003$ and $\frac{E_\infty}{E_0} = 10$.

Finally, Fig. 4 investigates the frictional response of the contact. In the presence of viscoelastic hysteresis, the asymmetric contact distribution developed within the contact area yields to friction coefficient:

$$\mu = -\frac{1}{\lambda p_\infty} \int_{-x_1}^{x_2} u'(x) p(x) dx \qquad (13)$$

where $u'(x)$ is the derivative of the surface displacement and $p_\infty$ the remote pressure. Notably, $\mu$ reflects viscoelastic energy dissipation occurring within the whole solid during sliding motion. Besides viscoelastic hysteresis due to cyclic deformations involving the bulk material, which completely characterizes friction in adhesiveless condition, a large additional frictional contribution is induced by viscoelastic adhesion hysteresis, roughly quantifiable through the adhesive friction coefficient:

$$\mu_a = \frac{G_1 - G_2}{\lambda p_\infty} \qquad (14)$$

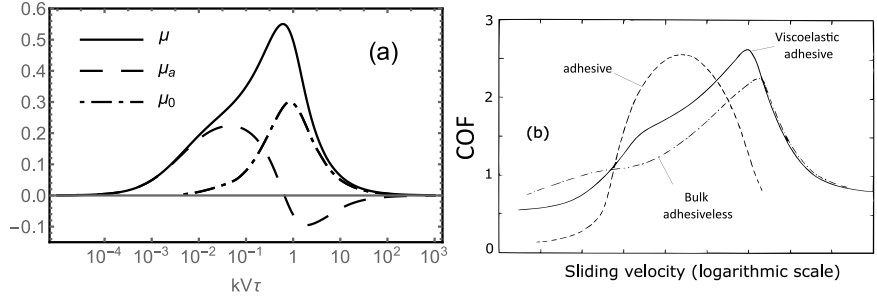

**Fig. 4.** (a) The predicted frictional behavior investigated as function of the dimensionless sliding velocity $kV\tau$. Solid, dashed and dot-dashed line refers respectively to overall friction coefficient $\mu$, adhesive friction coefficient $\mu_a$ (defined by Eq.(14)) and friction coefficient calculated in adhesiveless conditions $\mu_0$. Data refer to dimensionless energy of adhesion $\tilde{\gamma} = \frac{k\Delta\gamma}{\pi E_0^*} = 0.003$, dimensionless remote pressure $\tilde{p}_\infty = \frac{2p_\infty}{E_0^* k\Lambda} = 0.11$ and $\frac{E_\infty}{E_0} = 10$. (b) Experimental friction coefficient measured as function of sliding velocity from [20] with solid, dashed and dot-dashed line respectively referring to rubber sliding on smooth glass, clean silicon carbide, dusted silicon carbide.

$\mu$ and $\mu_a$ are plotted against the dimensionless sliding velocity. The friction coefficient developed in the adhesiveless contact case is also shown as $\mu_0$. The figure clearly highlights that adhesion hysteresis governs the frictional response at low velocity values (small-scale viscoelastic regime), where dissipation involving the bulk vanishes leading $\mu \sim \mu_a$. Adhesion still plays a major role at higher velocity, where the coupling between small- and large- scale viscoelasticity leads to significantly higher friction compared to the adhesiveless case. This phenomenon eventually results in $\mu \gg \mu_a + \mu_0$, suggesting that models based on summing up independent estimations of adhesive and bulk contributions do not adequately quantify the overall friction. This is indeed confirmed by Grosch's experiments conducted on sliding rubber [20], with data reported in Fig. 4b. The solid line refers to friction measured on a clean sliding rough surface, thus involving bulk viscoelasticity and adhesion, showing a similar trend qualitatively compared to our findings. The adhesive friction estimate is reported with dashed line, referring to measurement conducted on a smooth glass surface ensuring negligible bulk hysteresis. The adhesiveless condition (dot-dashed line) is instead ensured by means of a dusted silicon carbide surface.

## 4    Conclusion

In this paper, we present our recent study on sliding adhesive contact between viscoelastic solids and rigid substrates. The unknown contact domain is calculated by means of a novel energy approach [23,24], properly considering the non-conservative nature of the viscoelastic material. We investigate tribological features as function of the sliding velocity: the pull-off load, the contact area's length and the energy release rate at the contact trailing and leading edge exhibit non-monotonic trends ascribable to specific scale-dependent phenomena. Viscoelastic hysteretic losses localized at the edges of the contact make the system develop highly enhanced adhesive performance already at low velocity values, in agreement with experiments conducted on rigid cylinder rolling in adhesive contact with rubber substrates [14-16]. Our friction vs. sliding velocity trend is very similar to Grosch's experimental findings [20], reflecting the complex coupling between small- and large- scale viscoelasticity. Notably, summing up adhesive friction and bulk adhesiveless friction do not provide a correct estimation of the overall friction coefficient.